\documentclass[twocolumn,prb,superscriptaddress,amsmath]{revtex4}

\usepackage{graphicx}

\begin{document}

\title
{Probing quantum phases of ultracold atoms in optical lattices by
transmission spectra in cavity QED}
\date{\today}

\author{Igor B. Mekhov}
\email{Igor.Mekhov@uibk.ac.at} \affiliation{Institute for
Theoretical Physics, University of Innsbruck, Technikerstr. 25, 6020
Innsbruck, Austria} \affiliation{St. Petersburg State University, V.
A. Fock Institute of Physics, Ulianovskaya 1, 198504 St. Petersburg,
Russia}
\author{Christoph Maschler}
\author{Helmut Ritsch}
\email{Helmut.Ritsch@uibk.ac.at} \affiliation{Institute for
Theoretical Physics, University of Innsbruck, Technikerstr. 25, 6020
Innsbruck, Austria}

\begin{abstract}
\end{abstract}

\maketitle

Studies of ultracold gases in optical lattices~\cite{BlochNatPhys}
link many disciplines. They allow testing fundamental quantum
many-body concepts of condensed-matter physics in well controllable
atomic systems~\cite{BlochNatPhys}, e.g., strongly correlated
phases, quantum information processing. Standard methods to observe
quantum properties of Bose-Einstein condensates (BEC) are based on
matter-wave interference between atoms released from
traps~\cite{BlochNature,Lukin,Stoferle,Gritsev,Schellekens},
destroying the system. Here we propose a new, nondestructive in atom
numbers, method based on optical measurements, proving that atomic
quantum statistics can be mapped on transmission spectra of high-Q
cavities, where atoms create a quantum refractive index. This can be
extremely useful for studying phase transitions~\cite{Jaksch}, e.g.
between Mott insulator and superfluid states, since various phases
show qualitatively distinct light scattering. Joining the paradigms
of cavity quantum electrodynamics (QED) and ultracold gases will
enable conceptually new investigations of both light and matter at
ultimate quantum levels. We predict effects accessible in
experiments, which only recently became possible~\cite{Esslinger}.

All-optical methods to characterize atomic quantum statistics were
proposed for homogeneous BEC~\cite{You94,Jav94,You95,Jav95,Parkins}
and some modified spectral properties induced by BEC's were
attributed to collective emission~\cite{You94,Jav94}, recoil
shifts~\cite{Jav95} or local field effects~\cite{Morice}.

We show a completely different phenomenon directly reflecting atom
quantum statistics due to state-dependent dispersion. More
precisely, the dispersion shift of a cavity mode depends on the atom
number. If the atom number in some lattice region fluctuates from
realization to realization, the modes get a fluctuating frequency
shift. Thus, in the cavity transmission-spectrum, resonances appear
at different frequencies directly reflecting the atom number
distribution function. Such a measurement allows then to calculate
atomic statistical quantities, e.g., mean value and variance
reflected by spectral characteristics such as the central frequency
and width.

Different phases of a degenerate gas possess similar mean-field
densities but different quantum amplitudes. This leads to a
superposition of different transmission spectra, which e.g. for a
superfluid state (SF) consist of numerous peaks reflecting the
discreteness of the matter-field. Analogous discrete spectra
reversing the role of atoms and light, thus reflecting the photon
structure of electromagnetic fields, were obtained in cavity QED
with Rydberg atoms~\cite{Haroche} and solid-state superconducting
circuits~\cite{Schoelkopf}. A quantum phase transition towards a
Mott insulator state (MI) is characterized by a reduction of the
number of peaks towards a single resonance, because atom number
fluctuations are significantly suppressed~\cite{Gerbier,Campbell}.
As our detection scheme is based on nonresonant dispersive
interaction independent of a particular level structure, it can be
also applied to molecules~\cite{Volz,Winkler}.

\begin{figure}
\scalebox{0.6}[0.6]{\includegraphics{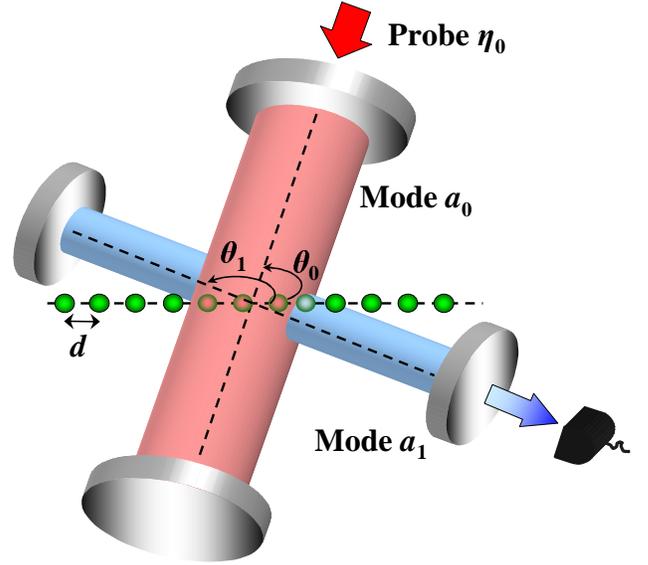}}
\caption{\label{fig1}
{\bf Schematic setup.} Atoms are periodically trapped in an optical
lattice created by laser beams, which are not shown in this figure.
Additionally, the atoms are illuminated by two light modes at the
angles $\theta_{0,1}$ with respect to the lattice axis.}
\end{figure}

We consider the quantized motion of $N$ two-level atoms in a deep
periodic optical lattice with $M$ sites formed by far off-resonance
standing wave laser beams~\cite{BlochNatPhys}. A region of $K\le M$
sites is coupled to two quantized light modes whose geometries (i.e.
axis directions or wavelengths) can be varied. This is shown in
Fig.~\ref{fig1} depicting two cavities crossed by a 1D string of
atoms in equally separated wells generated by the lattice lasers
(not shown). In practice two different modes of the same cavity
would do as well.

As shown in the Methods section, the Heisenberg equations for the
annihilation operators of two light modes $a_l$ ($l=0,1$) with
eigenfrequencies $\omega_l$ and spatial mode functions $u_l({\bf
r})$ are

\begin{eqnarray}\label{2}
\dot{a}_l= -i\left(\omega_l +\delta_l\hat{D}_{ll}\right)a_l
-i\delta_m\hat{D}_{lm}a_m -\kappa a_l+\eta_l(t), \\
\text{with} \quad \hat{D}_{lm}\equiv \sum_{i=1}^K{u_l^*({\bf
r}_i)u_m({\bf r}_i)\hat{n}_i},\nonumber
\end{eqnarray}
where $l\ne m$, $\delta_l=g^2/\Delta_{la}$, $g$ is the atom-light
coupling constant, $\Delta_{la} = \omega_l -\omega_a$ are the large
cavity-atom detunings, $\kappa$ is the cavity relaxation rate,
$\eta_l(t)=\eta_l e^{-i\omega_{lp}t}$ gives the external probe and
$\hat{n}_i$ are the atom number operators at a site with coordinate
${\bf r}_i$. We also introduce the operator of the atom number at
illuminated sites $\hat{N}_K=\sum_{i=1}^K{\hat{n}_i}$.

In a classical limit, Eq.~(\ref{2}) corresponds to Maxwell's
equations with the dispersion-induced frequency shifts of cavity
modes $\delta_l\hat{D}_{ll}$ and the coupling coefficient between
them $\delta_1\hat{D}_{10}$. For a quantum gas those quantities are
operators, which will lead to striking results: atom number
fluctuations will be directly reflected in such measurable
frequency-dependent observables. Thus, cavity transmission-spectra
will reflect atomic statistics.

Eq.~(\ref{2}) allows to express the light operators $a_l$ as a
function $f(\hat{n}_1,...,\hat{n}_M)$ of atomic occupation number
operators and calculate their expectation values for prescribed
atomic states $|\Psi\rangle$. We start with the well known examples
of MI and SF states and generalize to any $|\Psi\rangle$ later.

From the viewpoint of light scattering, the MI state behaves almost
classically as, for negligible tunneling, precisely
$\langle\hat{n}_i\rangle_\text{MI}=q_i$ atoms are well localized at
the $i$th site with no number fluctuations. It is represented by a
product of Fock states, i.e. $|\Psi\rangle_\text{MI}=\prod_{i=1}^M
|q_i\rangle_i\equiv |q_1,...,q_M\rangle$, with expectation values
\begin{eqnarray}\label{3}
\langle f(\hat{n}_1,...,\hat{n}_M)\rangle_\text{MI}=f(q_1,...,q_M),
\end{eqnarray}
since $\hat{n}_i|q_1,...,q_M\rangle=q_i|q_1,...,q_M\rangle$. For
simplicity we consider equal average densities
$\langle\hat{n}_i\rangle_\text{MI}=N/M\equiv n$
($\langle\hat{N}_K\rangle_\text{MI}=nK\equiv N_K$).

In our second example, SF state, each atom is delocalized over all
sites leading to local number fluctuations at a lattice region with
$K<M$ sites. Mathematically it is a superposition of Fock states
corresponding to all possible distributions of $N$ atoms at $M$
sites: $|\Psi\rangle_\text{SF}
=\sum_{q_1,...,q_M}\sqrt{N!/M^N}/\sqrt{q_1!...q_M!}
|q_1,...,q_M\rangle$. Although its average density
$\langle\hat{n}_i\rangle_\text{SF}=N/M$ is identical to a MI, it
creates different light transmission spectra. Expectation values of
light operators can be calculated from
\begin{eqnarray}\label{4}
\langle f(\hat{n}_1,...,\hat{n}_M)\rangle_\text{SF}=\frac{1}{M^N}
\sum_{q_1,...,q_M}\frac{N!} {q_1!...q_M!}f(q_1,...,q_M),
\end{eqnarray}
representing a sum of all possible ``classical'' terms. Thus, all
these distributions contribute to scattering from a SF, which is
obviously different from $\langle
f(\hat{n}_1,...,\hat{n}_M)\rangle_\text{MI}$ (\ref{3}) with only a
single contributing term.

\begin{figure}
\scalebox{0.9}[0.9]{\includegraphics{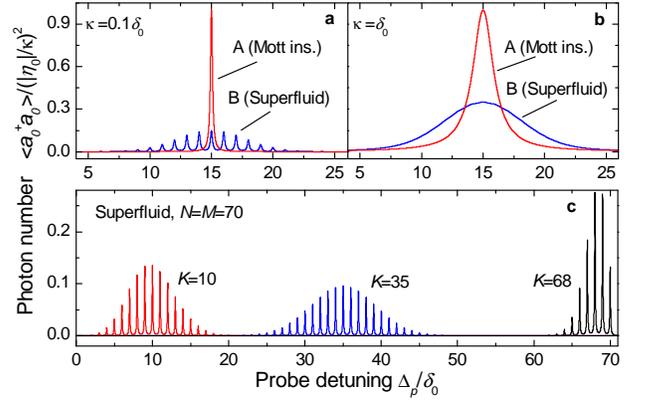}}
\caption{\label{fig2}
{\bf Photon number in a single cavity mode.} {\bf a,} Single
Lorentzian for MI (curve A) reflects non-fluctuating atom number.
Many Lorentzians for SF (curve B) reflect atom number fluctuations,
which are imprinted on the positions of narrow resonances. Here
$\kappa$ is smaller than satellite separation $\delta_0$
($\kappa=0.1\delta_0$), $N=M=30$, $K=15$. {\bf b,} The same as in
{\bf a} but $\kappa=\delta_0$ gives smooth broadened contour for SF.
{\bf c,} Spectra for SF with $N=M=70$ and different number of sites
illuminated $K=10,35,68$. The transmission spectra have different
forms, since different atom distribution functions correspond to
different $K$. $\kappa=0.05\delta_0$.}
\end{figure}

In the simple case of only one mode $a_0$ ($a_1\equiv 0$), the
stationary solution of Eq.~(\ref{2}) for the photon number reads
\begin{eqnarray}\label{5}
a^\dag_0a_0=f(\hat{n}_1,...,\hat{n}_M)=
\frac{|\eta_0|^2}{(\Delta_p-\delta_0\hat{D}_{00})^2 +\kappa^2},
\end{eqnarray}
where $\Delta_p=\omega_{0p}-\omega_0$ is the probe-cavity detuning.
We present transmission spectra in Fig.~\ref{fig2} for the case,
where $|u_{0}({\bf r}_i)|^2=1$, and $\hat{D}_{00}=\sum_{i=1}^K
\hat{n}_i$ reduces to $\hat{N}_K$. For a 1D lattice (see
Fig.~\ref{fig1}), this occurs for a traveling wave at any angle, and
standing wave transverse ($\theta_0=\pi/2$) or parallel
($\theta_0=0$) to the lattice with atoms trapped at field maxima.

For MI, the averaging of Eq.~(\ref{5}) according to Eq.~(\ref{3})
gives the photon number $\langle a^\dag_0a_0\rangle_\text{MI}$, as a
function of the detuning, as a single Lorentzian described by
Eq.~(\ref{5}) with width $\kappa$ and frequency shift given by
$\delta_0\langle\hat{D}_{00}\rangle_\text{MI}$ (equal to
$\delta_0N_K$ in Fig.~\ref{fig2}). Thus, for MI, the spectrum
reproduces a simple classical result of a Lorentzian shifted due to
dispersion.

In contrast, for a SF, the averaging procedure of Eq.~(\ref{4})
gives a sum of Lorentzians with different dispersion shifts
corresponding to all atomic distributions $|q_1,...,q_K\rangle$. So,
if each Lorentzian is resolved, one can measure a comb-like
structure by scanning the detuning $\Delta_p$. In Figs.~\ref{fig2}a
and \ref{fig2}c, different shifts of the Lorentzians correspond to
different possible atom numbers at $K$ sites (which due to atom
number fluctuations in SF, can take all values 0,1,2,...,$N$). The
Lorentzians are separated by $\delta_0$. Thus, we see that atom
number fluctuations lead to the fluctuating mode shift, and hence to
multiple resonances in the spectrum. For larger $\kappa$ the
spectrum becomes continuous (Fig.~\ref{fig2}b), but broader than
that for MI.

Scattering of weak fields does not change the atom number
distribution. However, as the SF is a superposition of different
atom numbers in a region with $K$ sites, a measurement projects the
state into a subspace with fixed $N_K$ in this region, and a
subsequent measurement on a time scale short to tunneling between
sites will yield the same result. One recovers the full spectrum of
Fig.~\ref{fig2} by repeating the experiment or with sufficient delay
to allow for redistribution via tunneling. Such measurements will
allow a time dependent study of tunneling and buildup of long-range
order. Alternatively, one can continue measurements on the reduced
subspace after changing a lattice region or light geometry.

\begin{figure}
\scalebox{0.9}[0.9]{\includegraphics{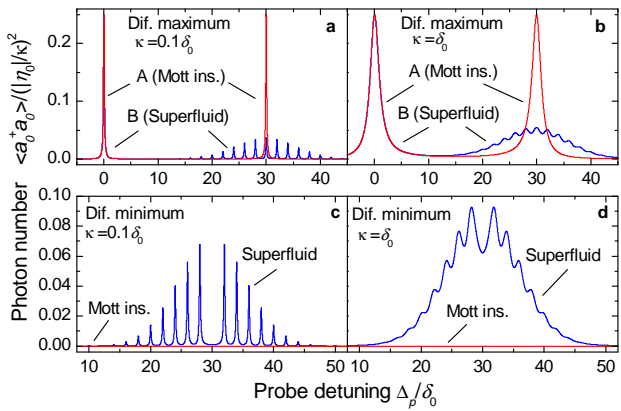}}
\caption{\label{fig3}
{\bf Photon number in one of two strongly coupled modes.} {\bf a,}
Diffraction maximum, doublet for MI (curve A) and spectrum with
structured right satellite for SF (curve B). Structure in the
satellite reflects atom number fluctuations in SF, while narrow
spectrum for MI demonstrates vanishing fluctuations. Here $\kappa$
is smaller than satellite separation $2\delta_0$
($\kappa=0.1\delta_0$), $K=15$. {\bf b,} The same as in {\bf a} but
$\kappa=\delta_0$ gives broadened satellite for SF. {\bf c,}
Diffraction minimum, zero field for MI and structured spectrum for
SF. Nonzero structured spectrum for SF reflects fluctuating
difference between atom numbers at odd and even sites, which exists
even for the whole lattice illuminated, $K=M$. Here $\kappa$ is
smaller than satellite separation $2\delta_0$
($\kappa=0.1\delta_0$), $K=30$. {\bf d,} The same as in {\bf c} but
$\kappa=\delta_0$ gives broadened contour for SF. $N=M=30$ in all
figures.}
\end{figure}

We now consider two modes with $\omega_0=\omega_1$, the probe
injected only into $a_0$ (Fig.~\ref{fig1}) and the mentioned
geometries where $\hat{D}_{00}=\hat{D}_{11}=\hat{N}_K$ (see
Fig.~\ref{fig3}). From Eq.~(\ref{2}), the stationary photon number
$a^\dag_1a_1=f(\hat{n}_1,...,\hat{n}_M)$ is
\begin{eqnarray}\label{6}
a^\dag_1a_1=\frac{\delta_1^2\hat{D}^\dag_{10}\hat{D}_{10}|\eta_0|^2}
{[\hat{\Delta}'^2_p-
\delta_1^2\hat{D}^\dag_{10}\hat{D}_{10}-\kappa^2]^2+4\kappa^2\hat{\Delta}'^2_p},
\end{eqnarray}
where $\hat{\Delta}'_p=\Delta_p-\delta_1\hat{D}_{11}$.

In a classical (and MI) case, Eq.~(\ref{3}) gives a two-satellite
contour (\ref{6}) reflecting normal-mode splitting of two
oscillators $\langle a_{0,1}\rangle$ coupled through atoms. This was
recently observed~\cite{Klinner} for collective strong coupling,
i.e., the splitting $\delta_1\langle\hat{D}_{10}\rangle$ exceeding
$\kappa$. The splitting depends on the geometry (see Eq.~(\ref{2}))
representing diffraction of one mode into another. Thus, our results
can be treated as scattering from a ``quantum diffraction grating''
generalizing Bragg scattering, well-known in different disciplines.
In diffraction maxima (i.e. $u_1^*({\bf r}_i)u_0({\bf r}_i)=1$) one
finds $\hat{D}_{10}=\hat{N}_K$ providing the maximal classical
splitting. In diffraction minima, one finds
$\hat{D}_{10}=\sum_{i=1}^K(-1)^{i+1}\hat{n}_i$ providing both the
classical splitting and photon number are almost zero.

In SF, Eq.~(\ref{4}) shows that $\langle
a^\dag_1a_1\rangle_\text{SF}$ is given by a sum of all classical
terms with all possible normal mode splittings. In a diffraction
maximum (Figs.~\ref{fig3}a,b), the right satellite is split into
components corresponding to all possible $N_K$ or extremely
broadened. In a minimum (Figs.~\ref{fig3}c,d), the splittings are
determined by all differences between atom numbers at odd and even
sites $\sum_{i=1}^K(-1)^{i+1}q_i$. Note that there is no classical
description of the spectra in a minimum, since here the classical
field (and $\langle a^\dag_1a_1\rangle_\text{MI}$) are simply zero
for any $\Delta_p$. Thus, for two cavities coupled at diffraction
minimum, the difference between the SF and MI states is even more
striking: one has a structured spectrum instead of zero signal.
Moreover, the difference between atom numbers at odd and even sites
fluctuates even for the whole lattice illuminated, giving nontrivial
spectra even for $K=M$.

In each of the examples in Figs.~\ref{fig2} and \ref{fig3}, the
photon number depends only on one statistical quantity, now called
$q$, $f(q_1,...,q_M)=f(q)$. For the single mode and two modes in a
maximum, $q$ is the atom number at $K$ sites. For two modes in a
minimum, $q$ is the atom number at odd (or even) sites. Therefore,
expectation values for some state $|\Psi\rangle$ can be reduced to
$\langle f\rangle_\Psi= \sum_{q=0}^Nf(q)p_\Psi (q)$, where $p_\Psi
(q)$ is the distribution function of $q$ in this state.

In high-Q cavities ($\kappa \ll \delta_0=g^2/\Delta_{0a}$), $f(q)$
is given by a narrow Lorentzian of width $\kappa$ peaked at some
frequency proportional to $q$ ($q=0,1,...,N$). The Lorentzian hight
is $q$-independent. Thus, $\langle f\rangle_\Psi$ as a function of
$\Delta_p$ represents a comb of Lorentzians with the amplitudes
simply proportional to $p_\Psi (q)$.

This is our central result. It states that the transmission spectrum
of a high-Q cavity $\langle a^\dag a (\Delta_p)\rangle_\Psi$
directly maps the distribution function of ultracold atoms $p_\Psi
(q)$, e.g., distribution function of atom number at $K$ sites.
Various atomic statistical quantities characterizing a particular
state can be then calculated: mean value (given by the spectrum
center), variance (determined by the spectral width) and higher
moments. Furthermore, transitions between different states will be
reflected in spectral changes. Deviations from idealized MI and SF
states~\cite{Lewenstein} are also measurable.

For SF, using $p_\text{SF}(q)$ (see Methods), we can write the
envelopes of the comb of Lorentzians shown in Figs.~\ref{fig2}a,c
and \ref{fig3}a,c. As known, the atom number at $K$ sites fluctuates
in SF with the variance $(\Delta N_K)^2=N_K(1-K/M)$. For example,
Fig.~\ref{fig2}c shows spectra for different lattice regions
demonstrating Gaussian and Poissonian distributions with the
spectral width $\sigma_\omega=\delta_0\sqrt{(\Delta N_K)^2}$,
directly reflecting the atom distribution functions in SF. For
$K\approx M$ the spectrum narrows, and, for the whole lattice
illuminated, shrinks to a single Lorentzian as in MI.

The condition $\kappa < \delta_0=g^2/\Delta_{0a}$ is already met in
present experiments. In the recent work~\cite{Esslinger}, where
setups of cavity QED and ultracold gases were joined to probe
quantum statistics of an atom laser with $^{87}$Rb atoms, the
parameters are $(g,\Delta_{0a},\kappa)=2\pi\times(10.4,30,1.4)$ MHz.
The setups of cavity cooling~\cite{Rempe,Kimble} are also very
promising.

For bad cavities ($\kappa\gg \delta_0=g^2/\Delta_{0a}$), the sums
can be replaced by integrals. The broad spectra in Figs.~\ref{fig2}b
and \ref{fig3}b,d are then given by convolutions of $p_\Psi(q)$ and
Lorentzians. For example, curve B in Fig.~\ref{fig2}b represents a
Voigt contour, well-know in spectroscopy of hot gases. Here, the
``inhomogeneous broadening'' is a striking contribution of quantum
statistics.

In summary, we exhibited that transmission spectra of cavities
around a degenerate gas in an optical lattice are distinct for
different quantum phases of even equal densities. Similar
information is also contained in the field amplitudes $\langle
a_{0,1}\rangle$ contrasting previous suggestions~\cite{Parkins} that
$\langle a_{0,1}\rangle$ probes only the average density. This
reflects (i) the orthogonality of Fock states corresponding to
different atom distributions and (ii) the different frequency shifts
of light fields entangled to those states. In general also other
optical phenomena and quantities depending nonlinearly on atom
number operators should similarly reflect the underlying quantum
statistics~\cite{We,ICAP,Meystre}.

\section*{Methods}
\subsection*{Derivation of Heisenberg equations}

A manybody Hamiltonian for our system presented in Fig.~\ref{fig1}
is given by
\begin{eqnarray}
H=\sum_{l=0,1}\hbar{\omega_l a^\dag_l a_l} +\int{d^3{\bf
r}\Psi^\dag({\bf r})H_{a1}\Psi({\bf r})}, {\rm with} \nonumber\\
 H_{a1}=\frac{{\bf
p}^2}{2m_a}+V_{\text {cl}}({\bf r})+\hbar
g^2\sum_{l,m=0,1}{\frac{u_l^*({\bf r})u_m({\bf r}) a^\dag_l
a_m}{\Delta_{ma}}},\nonumber
\end{eqnarray}
where $a_{0,1}$ are the annihilation operators of the modes of
frequencies $\omega_{0,1}$, wave vectors ${\bf k}_{0,1}$, and mode
functions $u_{0,1}({\bf r})$; $\Psi({\bf r})$ is the atom-field
operator. In the effective single-atom Hamiltonian $H_{a1}$, ${\bf
p}$ and ${\bf r}$ are the momentum and position operators of an atom
of mass $m_a$ trapped in the classical potential $V_{\text
{cl}}({\bf r})$, and $g$ is the atom--light coupling constant. We
consider off-resonant scattering where the detunings between fields
and atomic transition $\Delta_{la} = \omega_l -\omega_a$ are larger
than the spontaneous emission rate and Rabi frequencies. Thus, in
$H_{a1}$ the adiabatic elimination of the upper state, assuming
linear dipoles with adiabatically following polarization, was used.

For a one-dimensional lattice with period $d$ and atoms trapped at
$x_j=jd$ ($j=1,2,\hdots,M$) the mode functions are $u_{0,1}({\bf
r}_j)=\exp (ijk_{0,1x}d+i\phi)$ for traveling and $u_{0,1}({\bf
r}_j)=\cos (jk_{0,1x}d+\phi)$ standing waves with $k_{0,1x}=|{\bf
k}_{0,1}|\cos\theta_{0,1}$, $\theta_{0,1}$ are angles between the
mode and lattice axes, $\phi$ is some spatial phase shift (cf.
Fig.~\ref{fig1}).

Assuming the modes $a_{0,1}$ much weaker than the trapping beam, we
expand $\Psi({\bf r})$ using localized Wannier
functions~\cite{Jaksch} corresponding to the potential $V_{\text
{cl}}({\bf r})$ and keep only the lowest vibrational state at each
site (we consider a quantum degenerate gas): $\Psi({\bf
r})=\sum_{i=1}^{M}{b_i w({\bf r}-{\bf r}_i)}$, where $b_i$ is the
annihilation operator of an atom at site $i$ at a position ${\bf
r}_i$. Substituting this expansion in the Hamiltonian $H$, one can
get a generalized Bose-Hubbard model~\cite{Jaksch} including light
scattering. In contrast to ``Bragg spectroscopy'', which involves
scattering of matter waves~\cite{Stoferle}, and our previous
work~\cite{Maschler}, we neglect lattice excitations here and focus
on light scattering from atoms in some prescribed quantum states.

Neglecting atomic tunneling, the Hamiltonian reads:

\begin{eqnarray}
H=\sum_{l=0,1}{\hbar\omega_l a^\dag_l a_l}+ \hbar g^2
\sum_{l,m=0,1}{\frac{a^\dag_l
a_m}{\Delta_{ma}}}\left(\sum_{i=1}^K{J_{i,i}^{lm}\hat{n}_i}\right),
\nonumber
\end{eqnarray}
where $\hat{n}_i=b_i^\dag b_i$. For deep lattices the coefficients
$J_{i,i}^{lm}=\int{d{\bf r}}w^2({\bf r}-{\bf r}_i) u_l^*({\bf
r})u_m({\bf r})$ reduce to $J_{i,i}^{lm}=u_l^*({\bf r}_i)u_m({\bf
r}_i)$ neglecting spreading of atoms, which can be characterized
even by classical scattering~\cite{Slama}. The Heisenberg equations
obtained from this Hamiltonian are given by Eq.~(\ref{2}), were we
have added a relaxation term. Strictly speaking, a Langevin noise
term should be also added to Eq.~(\ref{2}). However, for typical
conditions its influence on the expectation values of normal ordered
field operators is negligible (see e.g.~\cite{Davidovich}). In this
paper, we are interested in the number of photons $\langle a_l^\dag
a_l\rangle$ only, which is a normal ordered quantity. Thus, one can
simply omit the noise term in Eq.~(\ref{2}).

\subsection*{Simple expressions for spectral line shapes in SF state}

We will now derive expressions for the spectra presented in
Figs.~\ref{fig2} and \ref{fig3} demonstrating relations between
atomic quantum statistics and the transmission spectra for the SF
state. As has been mentioned in the main text, in all examples
presented in Figs.~\ref{fig2} and \ref{fig3}, the photon number
depends only on a single statistical quantity, which we denote as
$q$. Using this fact, the multinomial distribution in Eq.~(\ref{4})
reduces to a binomial, which can be directly derived from
Eq.~(\ref{4}): $\langle f\rangle_\text{SF}=
\sum_{q=0}^Nf(q)p_\text{SF}(q)$ with
$p_\text{SF}(q)=N!/[q!(N-q)!](Q/M)^q(1-Q/M)^{N-q}$ and a single sum
instead of $M$ ones. Here $Q$ is the number of specified sites: $Q$
is equal to $K$ for one mode and two modes in a maximum; $Q$ is the
number of odd (or even) sites for two modes in a minimum ($Q=M/2$
for even $M$). This approach can be used for other geometries, e.g.,
for two modes in a minimum and $K<M$, where Eq.~(\ref{4}) can be
reduced to a trinomial distribution.

As a next approximation we consider $N,M\gg1$, but finite $N/M$,
leading to the Gaussian distribution
$p_\text{SF}(q)=1/(\sqrt{2\pi}\sigma_q)\exp{[-(q-\tilde{q})^2/2\sigma_q^2]}$
with central value $\tilde{q}=NQ/M$ and width
$\sigma_q=\sqrt{N(Q/M)(1-Q/M)}$.

In high-Q cavities ($\kappa \ll \delta_0=g^2/\Delta_{0a}$), $f(q)$
is a narrow Lorentzian of width $\kappa$ peaked at some
$q$-dependent frequency, now called $\Delta_p^q$. Since the
Lorentzian hight is $q$-independent, $\langle f\rangle_\text{SF}$ as
a function of $\Delta_p$ is a comb of Lorentzians with the
amplitudes proportional to $p_\text{SF}(q)$.

Using the Gaussian distribution $p_\text{SF}(q)$,we can write the
envelope of such a comb. For a single mode [Fig.~\ref{fig2}a,c,
Eq.~(\ref{5})], we find $\Delta_p^q\approx \delta_0q$ with the
envelope
\begin{eqnarray}
\langle a_0^\dag a_0 (\Delta_p^q)\rangle_\text{SF}= \frac{\alpha
\delta_0}{\sqrt{2\pi}\sigma_\omega}
e^{-(\Delta_p^q-\tilde{\Delta}_p)^2/2\sigma_\omega^2},\nonumber
\end{eqnarray}
where the central frequency $\tilde{\Delta}_p=\delta_0N_K$, spectral
width $\sigma_\omega=\delta_0\sqrt{N_K(1-K/M)}$, and
$\alpha=|\eta_0|^2/\kappa^2$. So, the spectrum envelopes in
Fig.~\ref{fig2}a,c are well described by Gaussians of widths
strongly depending on $K$.

For $K\rightarrow 0$ and $K\rightarrow M$, the binomial distribution
$p_\text{SF}(q)$ is well approximated by a Poissonian distribution,
which is demonstrated in Fig.~\ref{fig2}c for $K=10$ and $K=68$. For
$K=M$ the spectrum shrinks to a single Lorenzian, since the total
atom number at $M$ sites does not fluctuate.

In other examples (Figs.~\ref{fig3}a and \ref{fig3}c), the above
expression is also valid, although with other parameters. For two
modes in a diffraction maximum (Fig.~\ref{fig3}a), the central
frequency, separation between Lorentzians and width are doubled:
$\tilde{\Delta}_p=2\delta_0N_K$, $\Delta_p^q\approx 2\delta_0q$ and
$\sigma_\omega=2\delta_0\sqrt{N_K(1-K/M)}$;
$\alpha=|\eta_0|^2/(2\kappa^2)$. The left satellite at $\Delta_p=0$
has a classical amplitude $|\eta_0|^2/(4\kappa^2)$.

The nonclassical spectrum for two waves in a diffraction minimum
(Fig.~\ref{fig3}c) is centered at $\tilde{\Delta}_p=\delta_0N$, with
components at $\Delta_p^q\approx 2\delta_0q$, and is very broad,
$\sigma_\omega=\delta_0\sqrt{N}$; $\alpha=|\eta_0|^2/\kappa^2$.

For bad cavities ($\kappa\gg \delta_0$), the sums can be replaced by
integrals with the same parameters $\tilde{\Delta}_p$ and
$\sigma_\omega$ as for $\kappa < \delta_0$. For a single mode,
Fig.~\ref{fig2}b represents a Voigt contour
\begin{eqnarray}
\langle a_0^\dag a_0 (\Delta_p)\rangle_\text{SF}=
\frac{|\eta_0|^2}{\sqrt{2\pi}\sigma_\omega}\int_0^\infty
\frac{e^{-(\omega-\tilde{\Delta}_p)^2/2\sigma_\omega^2}d\omega}{(\Delta_p-\omega)^2
+\kappa^2}.\nonumber
\end{eqnarray}
For two modes in a diffraction minimum the photon number
(Fig.~\ref{fig3}d) is
\begin{eqnarray}
\langle a_1^\dag a_1\rangle_\text{SF}=
\frac{|\eta_0|^2}{\sqrt{2\pi}\sigma_\omega}\int_{-\infty}^\infty
\frac{\omega^2e^{-\omega^2/2\sigma_\omega^2}d\omega}
{(\Delta'^2_p-\omega^2-\kappa^2)^2 +4\kappa^2\Delta'^2_p},\nonumber
\end{eqnarray}
where $\Delta'_p=\Delta_p-\tilde{\Delta}_p$, while in a maximum
(Fig.~\ref{fig3}b)
\begin{eqnarray}
\langle a_1^\dag a_1\rangle_\text{SF}=
\frac{|\eta_0|^2}{4\sqrt{2\pi}\sigma_\omega}\int_0^\infty
\frac{\omega^2e^{-(\omega-\tilde{\Delta}_p)^2/2\sigma_\omega^2}d\omega}
{[\Delta_p(\Delta_p-\omega)+\kappa^2]^2 +\kappa^2\omega^2}.\nonumber
\end{eqnarray}

\section*{Acknowledgments}
The work was supported by FWF (P17709 and S1512). While preparing
this manuscript, we became aware of a closely related research in
the group of P. Meystre. We are grateful to him for sending us the
preprint~\cite{Meystre} and stimulating discussions.

All authors equally contributed to the paper.

Correspondence and request for materials should be addressed to
I.B.M.

\section*{Competing financial interests}
The authors declare that they have no competing financial interests.


\begin{thebibliography}{30}
\bibitem{BlochNatPhys} Bloch, I. Ultracold quantum gases in optical
lattices. {\it Nat. Phys.} {\bf 1}, 23--30 (2005).
\bibitem{BlochNature} F{\"o}lling, S. {\it et al.} Spatial quantum noise
interferometry in expanding ultracold atom clouds. {\it Nature} {\bf
434}, 481--484 (2005).
\bibitem{Lukin} Altman, E., Demler, E., \& Lukin, M. D. Probing
many-body states of ultracold atoms via noise correlations. {\it
Phys.\ Rev.\ A} {\bf 70}, 013603 (2004).
\bibitem{Stoferle} St{\"o}ferle, T., Moritz, H., Schori, C.,
K{\"o}hl, M. \& Esslinger, T. Transition from a strongly interacting
1D superfluid to a Mott insulator. {\it Phys.\ Rev.\ Lett.} {\bf
92}, 130403 (2004).
\bibitem{Gritsev} Gritsev, V., Altman, E., Demler, E. \&
Polkovnikov, A. Full quantum distribution of contrast in
interference experiments between interacting one-dimensional Bose
liquids. {\it Nat. Phys.} {\bf 2}, 705--709 (2006).
\bibitem{Schellekens} Schellekens, M. {\it et al.} Hanbury Brown
Twiss effect for ultracold quantum gases. {\it Science} {\bf 310},
648--651 (2005).
\bibitem{Jaksch} Jaksch, D., Bruder, C., Cirac, J. I., Gardiner, C.
W. \& Zoller, P. Cold bosonic atoms in optical lattices {\it Phys.\
Rev.\ Lett.} {\bf 81}, 3108--3111 (1998).
\bibitem{Esslinger} Bourdel, T. {\it et al.} Cavity QED detection of
interfering matter waves. {\it Phys.\ Rev.\ A} {\bf 73}, 043602
(2006).
\bibitem{You94} You, L., Lewenstein, M. \& Cooper, J. Line shapes for light
scattered from Bose-Einstein condensates. {\it Phys.\ Rev.\ A} {\bf
50}, R3565--R3568 (1994).
\bibitem{Jav94} Javanainen, J. Optical signatures of a tightly confined Bose
condensate. {\it Phys.\ Rev.\ Lett.} {\bf 72}, 2375--2378 (1994).
\bibitem{You95} You, L., Lewenstein, M., \& Cooper, J. Quantum field theory
of atoms interacting with photons. II. Scattering of short laser
pulses from trapped bosonic atoms. {\it Phys.\ Rev.\ A} {\bf 51},
4712--4727 (1995).
\bibitem{Jav95} Javanainen, J. \& Ruostekoski, J. Off-resonance light
scattering from low-temperature Bose and Fermi gases. {\it Phys.\
Rev.\ A} {\bf 52}, 3033--3046 (1995).
\bibitem{Parkins} Parkins, A. S. \& Walls, D. F. The physics of trapped
dilute-gas Bose-Einstein condensates. {\it Phys. Rep.} {\bf 303},
1-80 (1998).
\bibitem{Morice} Morice, O., Castin, Y. \& Dalibard, J. Refractive index of a
dilute Bose gas. {\it Phys.\ Rev.\ A} {\bf 51}, 3896--3901 (1995).
\bibitem{Haroche} Brune, M., {\it et al.} Quantum Rabi
oscillation: a direct test of field quantization in a cavity. {\it
Phys.\ Rev.\ Lett.} {\bf 76}, 1800--1803 (1996).
\bibitem{Schoelkopf} Gambetta, J. {\it et al.} Qubit-photon interactions in
a cavity: Measurement-induced dephasing and number splitting. {\it
Phys.\ Rev.\ A} {\bf 74}, 042318 (2006).
\bibitem{Campbell} Campbell, G. K. {\it et al.} Imaging the Mott insulator shells
by using atomic clock shifts. {\it Science} {\bf 313}, 649--652
(2006).
\bibitem{Gerbier} Gerbier, F., F{\"o}lling, S., Widera, A., Mandel,
O. \& Bloch, I. Probing number squeezing of ultracold atoms across
the superfluid-Mott insulator transition. {\it Phys.\ Rev.\ Lett.}
{\bf 96}, 090401 (2006).
\bibitem{Volz} Volz, T. {\it et al.} Preparation of a quantum state with one
molecule at each site of an optical lattice. {\it Nat. Phys.} {\bf
2}, 692--695 (2006).
\bibitem{Winkler} Winkler, K. {\it et al.} Repulsively bound atom pairs in an optical
lattice. {\it Nature} {\bf 441}, 853--856 (2006).
\bibitem{Klinner} Klinner, J., Lindholdt, M., Nagorny, B. \& Hemmerich, A.
Normal mode splitting and mechanical effects of an optical lattice
in a ring cavity. {\it Phys.\ Rev.\ Lett.} {\bf 96}, 023002 (2006).
\bibitem{Lewenstein} Lewenstein, M. {\it et al.} Ultracold atomic gases in optical
lattices: Mimicking condensed matter physics and beyond.
cond-mat/0606771.
\bibitem{Rempe} Maunz, P. {\it et al.} Cavity cooling of a single
atom. {\it Nature} {\bf 428}, 50--52 (2004).
\bibitem{Kimble} Hood, C. J., Lynn, T. W., Doherty, A. C., Parkins,
A. S. \& Kimble, H. J. The atom-cavity microscope: single atoms
bound in orbit by single photons. {\it Science} {\bf 287},
1447--1453 (2000).
\bibitem{We} Mekhov, I. B., Maschler, C. \& Ritsch, H. Cavity
enhanced light scattering in optical lattices to probe atomic
quantum satistics. quant-ph/0610073, Phys. Rev. Lett. \textbf{98},
100402 (2007); Mekhov, I. B., Maschler, C. \& Ritsch, Light
scattering from ultracold atoms in optical lattices as an optical
probe of quantum statistics. quant-ph/0702193, Phys. Rev. A
\textbf{76}, 053618 (2007).
\bibitem{ICAP} Mekhov, I. B., Maschler, C. \& Ritsch, H., Light scattering
from atoms in an optical lattice: optical probe of quantum
statisticsin, {\it Books of abstracts for the XX International
Conference on Atomic Physics, ICAP, Innsbruck, 2006}, p. 309 and
conference web-site.
\bibitem{Meystre} Chen, W., Meiser, D. \& Meystre, P. Cavity QED
determination of atomic number statistics in optical lattices.
quant-ph/0610029.
\bibitem{Maschler} Maschler, C. \& Ritsch, H. Cold atom dynamics in a
quantum optical lattice potential. {\it Phys.\ Rev.\ Lett.} {\bf
95}, 260401 (2005); C. Maschler, I. B. Mekhov, and H. Ritsch.
Ultracold atoms in optical lattices generated by quantized light
fields. e-print arXiv:0710.4220.
\bibitem{Slama} Slama, S., von Cube, C., Kohler, M., Zimmermann, C.
\& Courteille, P. V. Multiple reflections and diffuse scattering in
Bragg scattering at optical lattices. {\it Phys.\ Rev.\ A} {\bf 73},
023424 (2006).
\bibitem{Davidovich} Davidovich, L. Sub-Poissonian processes in quantum
optics. {\it Rev.\ Mod.\ Phys.} {\bf 68}, 127 (1996).
\end{thebibliography}
\end{document}